\documentclass{article}

\usepackage{times}
\usepackage{graphicx}
\usepackage{authblk}

\usepackage[utf8]{inputenc}

\usepackage{setspace}
\usepackage{subfigure}

\usepackage{amsmath}
\usepackage{amsfonts}

\usepackage{ifpdf}

\ifpdf
\DeclareGraphicsExtensions{.pdf, .jpg, .tif}
\else
\DeclareGraphicsExtensions{.eps, .jpg}
\fi

\graphicspath{{./}{./figures/}}

\begin{document}


\title{Bayesian clustering in decomposable graphs}

\renewcommand\Affilfont{\small}
\author[1]{Luke Bornn}
\affil[1]{Department of Statistics, University of British Columbia.  l.bornn@stat.ubc.ca}

\author[2]{Fran\c{c}ois Caron}
\affil[2]{INRIA Bordeaux Sud-Ouest, Institut de Math\'ematiques de Bordeaux, University of Bordeaux. francois.caron@inria.fr}

\date{June 10, 2011}


\maketitle

\begin{abstract}
	In this paper we propose a class of prior distributions on decomposable graphs, allowing for improved modeling flexibility.  While existing methods solely penalize the number of edges, the proposed work empowers practitioners to control clustering, level of separation, and other features of the graph.  Emphasis is placed on a particular prior distribution which derives its motivation from the class of product partition models; the properties of this prior relative to existing priors is examined through theory and simulation.  We then demonstrate the use of graphical models in the field of agriculture, showing how the proposed prior distribution alleviates the inflexibility of previous approaches in properly modeling the interactions between the yield of different crop varieties.  Lastly, we explore American voting data, comparing the voting patterns amongst the states over the last century.
\end{abstract}

\section{Introduction}

This paper is concerned with the inference of the conditional independence graph $\mathcal{G}$ of a multivariate random vector $Y$ of dimension $n$, a problem sometimes referred to as structure learning. We focus here on undirected decomposable graphs, whose popularity is mainly due to the tractable factorization they allow for the likelihood (\cite{Dawid1993a, Lauritzen1996}); related work for directed graphical models can be found in \cite{Koller2009a}.  Learning the conditional independence graph $\mathcal{G}$ is an onerous task due to the large number of graphs on a set of $n$ nodes, or variables. It is possible using optimization methods to find the graph which best fits the data according to some metric~\cite{Meinshausen2006a,Yuan2007,Friedman2008}; alternatively Bayesian model averaging may be used to accommodate for uncertainty in the estimated graph, or maximum a posteriori estimation may be used to select a given model from the posterior over graphs. Such an approach relies on a prior distribution $\pi(\mathcal{G})$ over the set of decomposable graphs of a given size; through Bayes theorem, this prior is updated based on the data to give an a posteriori estimate of the distribution over graphs.

Current approaches have been limited in their ability to accommodate varying forms of prior information on the graph.  For instance, in an effort to encourage interpretable graphs, the standard approach has been to penalize the number of edges (conditional dependencies) in the graph.  However, many situations exist where one might expect variables to be clustered together and the graph to exhibit block structure. At the moment no such prior distribution exists to handle this problem.  Our contribution in this article is to propose a class of prior distributions motivated from the class of product partition models which will allow improved flexibility in the specification of prior information on the graph.

The field of agriculture is particularly suitable to the application of graphical models.  Due to large spatial domains as well as multifarious crop varieties, it is valuable to have models which both handle the complexity of the biophysical process as well as allow straightforward interpretation.  In particular, one might examine the set of zero/non-zero correlations between crop varieties' yields, using the presence or absence of edges to make decisions regarding crop management, marketing, and insurance policies.  In addition, due to small sample sizes in many agricultural applications, the choice of prior distribution becomes particularly important.


\section{Bayesian Inference on Decomposable Graphs}

We begin with a brief overview of graphical models, following the exposition in \cite{Dawid1993a}; see also \cite{Lauritzen1996} for further details on graphical models.  Let $\mathcal{G}=(V,E)$ be a graphical model with vertices $V=\{1,\ldots,n\}$ and pairwise edges $E$.  The pair of nodes $\{i,j\} \in V$ are adjacent if $(i,j) \in E$, and a subset $C \subset V$ is said to be complete if all its elements are adjacent to each other.  A complete subgraph that is maximal (i.e. not contained within another complete subgraph) is called a clique. An ordering of the cliques of an undirected graph, ($C_1,\ldots,C_{n_c}$) is said to be perfect if the vertices of each clique $C_i$ also contained in any previous clique $C_1,\ldots,C_{i-1}$ are all members of one previous clique; that is, for $i=2,3,\ldots,n_c$
$$
H_i=C_i\cap\cup_{j=1}^{i-1}C_j\subseteq C_h
$$
for some $h\in\{1,2,\ldots,i-1\}$. The sets $H_i$, $i=1,\ldots,n_c-1$ are called separators. We write $S_1,\ldots,S_{n_s}$ the non-empty separators (some might appear multiple times). If an undirected graph admits a perfect ordering it is said to be decomposable.

We associate to each vertex $i$ a random variable $Y_i$. For $A\subseteq V$, let $Y_A=\{Y_i|i\in A\}$. A distribution $P$ over $V$ is Markov with respect to $\mathcal{G}$ if, for any decomposition $(A,B)$ of $\mathcal{G}$, $X_A$ is independent of $X_B$ given $X_{A\cap B}$. The widespread use of decomposable models is due to the resulting factorization of densities. Specifically, if $P$ satisfies the conditional independencies implied by a decomposable graph $\mathcal{G}$, then the likelihood of the graphical model specified by $P$ can be factorized according to the graph's cliques and separators
\begin{equation}
p(y|\mathcal{G},\theta) = \frac{\prod_{i=1}^{n_c} p(y_{C_i}|\theta_{C_i})}{\prod_{j=1}^{n_s} p(y_{S_j}|\theta_{S_{j}})}\label{eq:likelihood}
\end{equation}
where $\theta$ is a quantity parameterizing the graphical model $P$ over the graph $\mathcal{G}$ and satisfying some consistency conditions with respect to $\mathcal{G}$ (\cite{Dawid1993a}).

Traditionally, focus has been on Gaussian graphical models, also known as covariance selection models (\cite{Dempster1972a}) where $P=N_n(\mu,\Sigma)$ is a $n$-dimensional multivariate Gaussian distribution and $\theta$ is the $n\times n$ covariance matrix $\Sigma$. Conditional independence structure is represented by the precision matrix $\Sigma^{-1}$.  If the edge  $(i,j) \notin E$, then the variables $Y_i$ and $Y_j$ are conditionally independent given the remaining variables, and $\Sigma^{-1}_{(i,j)} = \Sigma^{-1}_{(j,i)} = 0$.  As such, the Gaussian graphical model may be factorized as (\ref{eq:likelihood}) with the covariance $\Sigma$ replacing $\theta$, and the corresponding likelihood terms written as
\begin{align}
 p(y_B|\Sigma_B)=(2\pi)^{-|B|/2}\det(\Sigma_B)^{-|B|/2}\exp[-\frac{1}{2}tr(S_B(\Sigma_B)^{-1})]
\label{eq:sublikelihood}
\end{align}
for each complete set $B$, where $|B|$ denotes the cardinality of $B$ and $S_B$ is the empirical covariance matrix of $y_B$.

From a Bayesian perspective, we are interested in the posterior distribution $p(\theta,\mathcal{G}|y)\propto p(y|\theta,\mathcal{G})p(\theta|\mathcal{G})\pi(\mathcal{G})$. Much work has been dedicated to specifying proper priors $p(\theta|\mathcal{G})$, see e.g. (\cite{Giudici1999a, Dawid1993a}). The main focus of this paper is the specification of a prior distribution $\pi(\mathcal{G})$ over the space of decomposable graphs. As this space is very large compared to the number of observations, it is crucial to add as much prior information as possible on the structure of the unknown graph $\mathcal{G}$. Moreover, we are generally interested in obtaining sparse graph estimates for needs of interpretation and prediction. Up until now, the specification of $\pi(\mathcal{G})$ has been limited to the uniform distribution, or priors which penalize the complexity as measured by the number of edges.  This brings us to the focus of this work, namely a class of prior distributions $\pi(\mathcal{G})$ which subsumes control over the structure and features of $\mathcal{G}$.


\section{Priors on Decomposable Graphs}

\subsection{Previous work}

While early work on inference in decomposable models often assumed a uniform prior over graphs (i.e. \cite{Giudici1999a}), such priors put considerable mass on models of intermediate size.
In an effort to put more weight on smaller graphs, several authors have proposed using  a binomial prior distribution with parameter $\rho$ on the number of edges $r$ in the graph. This yields priors of the type~\cite{Dobra2004a,Jones2005a}%
\begin{equation}
\pi(\mathcal{G})\propto \rho^{r}(1-\rho)^{m-r}\label{eq:binomial}
\end{equation}
where $m=\frac{n(n-1)}{2}$ is the maximal number of possible edges on $n$ nodes. When
$\rho=1/2$, it reduces to the forementioned uniform prior over graphs.  \cite{Jones2005a} suggest the use of $\rho = 2/(n-1)$, motivated from the resulting density's peak at $n$ edges in the unconstrained graph.  Some authors also consider adding a hierarchical Beta prior $\rho \sim \text{\tt {B{\small e}}}(a,b)$ (\cite{Carvalho2009b}), giving the marginal prior on the graph as
\begin{equation*}
\pi(\mathcal{G}) = \int_0^1 \pi(\mathcal{G} | \rho) \pi(\rho) d\rho \propto \frac{\beta(a+k, b+m-k)}{\beta(a,b)}
\end{equation*}
where $\beta(\cdot,\cdot)$ is the beta function.  \cite{Carvalho2009b} suggest a default choice of $a=b=1$, implying a uniform prior on $\rho$.  Interestingly, the resulting prior on $\mathcal{G}$ is
\begin{equation*}
\pi(\mathcal{G}) = \frac{1}{m+1}  {m \choose r}^{-1},
\end{equation*}
which penalizes medium-sized graphs as desired.   Such a prior weights each graph according to the number of graphs in the unrestricted space with the same number of edges.  However, as shown by \cite{Armstrong2009}, the space of decomposable graphs can be considerably different than the unrestricted space. To address this, \cite{Armstrong2009} have proposed a uniform prior on decomposable graphs given the number of edges. However, calculating the number of decomposable graphs of a given size is an arduous task: there exists no list in the literature of decomposable graphs and their breakdown in terms of number of edges, nor are there straightforward ways of computing such quantities.  As a result, \cite{Armstrong2009} proposes an MCMC estimation scheme, testing its accuracy up to 12 nodes, although such a scheme will likely become prohibitive in higher dimensions.

\begin{figure}
  \centering
      \includegraphics[width=0.95\textwidth]{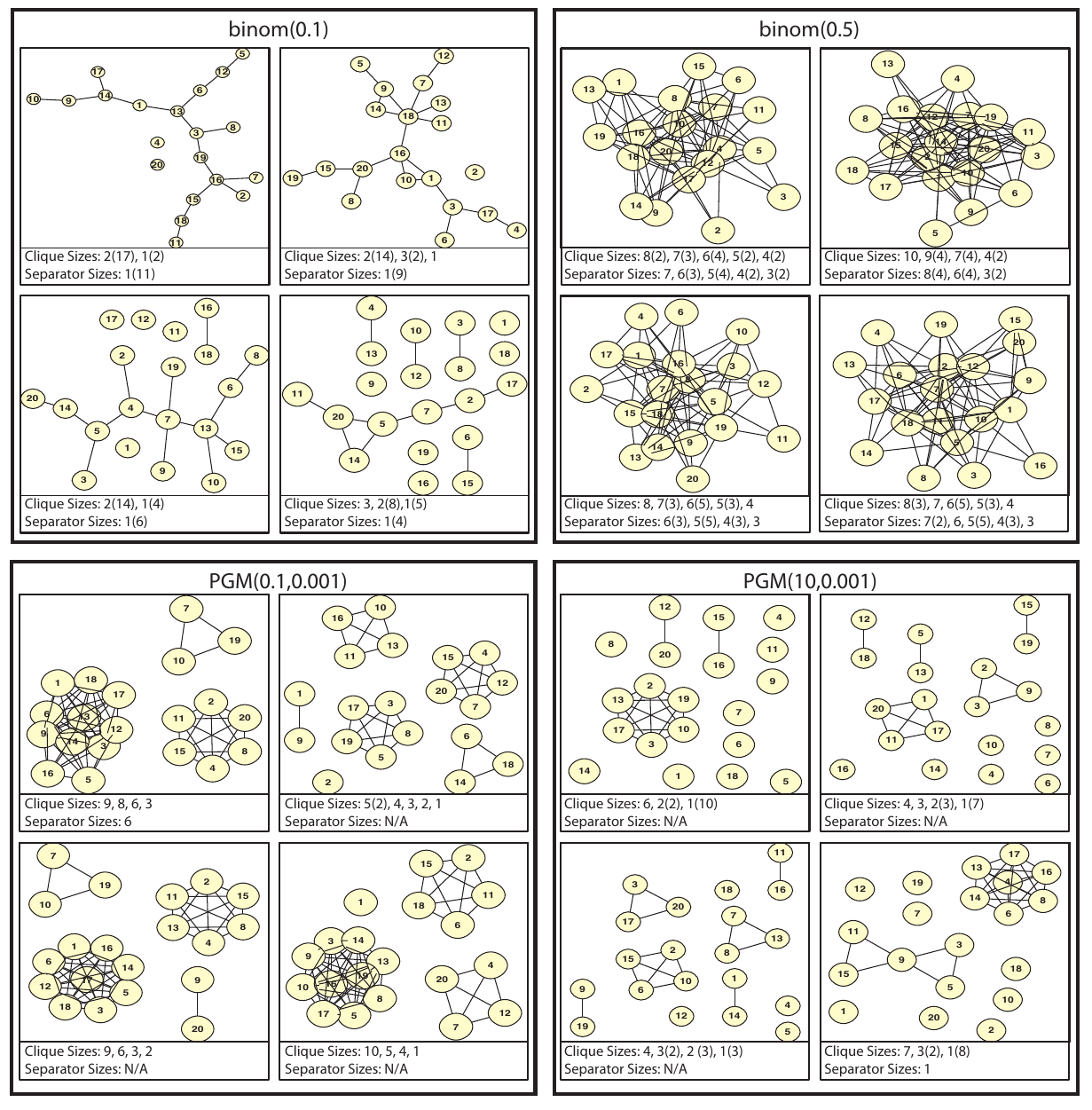}
  \caption{Four random samples from binomial and product graphical model (PGM) priors.  Clique and separator sizes for each graph are also shown (``Clique Sizes: $2(3)$'' implies 3 cliques of size 2).  $4$ million samples were generated using Markov chain Monte Carlo, and every millionth is shown.  While the binomial is characterized by large strings and many separators, the product graphical model allows one to induce clustering by setting $b$ small.}
	\label{fig:PriorSamples}
\end{figure}

While the priors in the above references allow one to control the size of the resulting graphs through the number of edges, often doing so results in undesirable graph structures, namely those with a high number of separators and long strings of nodes.  Figure \ref{fig:PriorSamples}(top)
shows random samples from a binomial prior over $20$-node graphs with $\rho = 0.1$ (closely echoing the choice of \cite{Jones2005a}, namely $\rho = 2/(n-1) \approx 0.1$) and $\rho = 0.5$ (the uniform prior).  We see from this plot that there is no clustering of the cliques, making interpretation difficult.  In addition, the long strings/trees seen for $\rho = 0.1$ do not mesh with reality in most cases. Clearly such a class of priors is not suitable if one suspects clustering amongst the variables, clique sizes to be upper (or lower) bounded, or nearly full separation between cliques.  Our focus therefore is on moving beyond priors which focus on the number of edges to priors which focus on graph (clique and separator) structure.

\subsection{A new prior distribution on decomposable graphs}

Motivated from the class of product partition models (\cite{Hartigan1990,Barry1992,Barry1993}), we consider prior distributions of the form
\begin{equation}
\pi(\mathcal{G})\propto\frac{\prod_{j=1}^{n_c}\psi_{C}(C_{j})}{\prod_{j=1}%
^{n_s}\psi_{S}(S_{j})} \label{eq:cohesion}
\end{equation}
where $\psi_{C}$ and $\psi_{S}$  are respectively called the clique/separator cohesion functions, with the convention that $\psi_{S}(\emptyset)=1$.  Evidently one could choose to penalize only cliques or separators by setting $\psi_{C}$ or $\psi_{S}$ to constant values. Alternatively, one could simply penalize clique sizes by setting $\psi_{B} = a |B|$.  Motivated from the class of product partition models, consider the cohesion functions $\psi_{C}(B)=a(|B|-1)!$ and $\psi_{S}(B)=\frac{1}{b}(|B|-1)!$, $a>0,$ $b>0$, hence
\begin{equation}
\pi(\mathcal{G})\propto a^{n_c}b^{n_s}\frac{\prod_{j=1}%
^{n_c}(|C_{j}|-1)!}{\prod_{j=1}^{n_s}(|S_{j}|-1)!}\label{eq:ppm_prior}
\end{equation}
The factorial terms result in predilection towards large cliques and small separators -- a desirable trait in terms of interpretability of the resulting graph.  For instance, even if $a=b=1$ with $20$ nodes, the completely connected graph would be preferred over the complete independence graph by a factor of $20!$.  The parameters $a$ and $b$ respectively tune the number of cliques and
separators in the decomposable graph. For $a$ small, the prior will favour a small number of large cliques.  Likewise for $b$, with small values favouring fewer separators.  Figure \ref{fig:PriorSamples} (bottom) shows samples from this prior.  Because of its relation to product partition models (described later), we term this prior the product graphical model prior.  To clearly demonstrate the control the product graphical model prior (\ref{eq:ppm_prior}) gives relative to the binomial prior, we set $b=1/1000$, highly penalizing the number of separators and hence resulting in highly separated cliques.  In addition, we look at two different values for $a$; $a=0.1$, resulting in fewer and larger cliques, and $a=10$, resulting in more (but smaller) cliques. Fig. \ref{fig:PriorSamples} demonstrates the ability of the prior to induce clustering of the cliques, and therefore sparsity in correlation.

\begin{figure}
  \centering
      \includegraphics[width=0.95\textwidth]{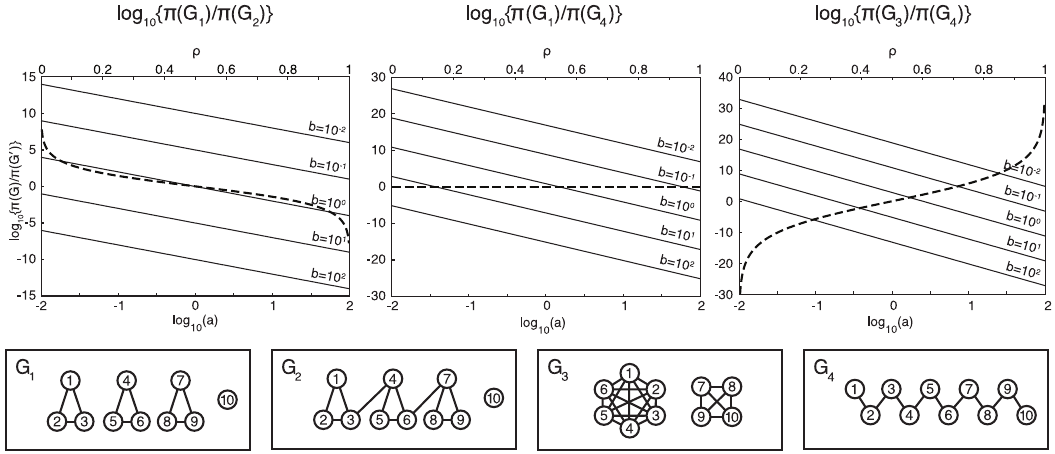}
  \caption{Log ratio of priors over two graphs for product graphical model prior for various $a$, $b$ (solid, bottom axis) and binomial prior for various $\rho$ (dashed, top axis).  While the binomial prior allows one to control the number of edges, for instance choosing $\mathcal{G}_1$ over $\mathcal{G}_2$, the same parameter would seldom choose $\mathcal{G}_3$ over $\mathcal{G}_4$, despite $\mathcal{G}_3$ having a sparse covariance matrix, and $\mathcal{G}_4$ having a saturated covariance matrix.}
	\label{fig:PriorRatios}
\end{figure}

We have seen some general properties of the prior (\ref{eq:ppm_prior}), namely the ability to control the number of cliques and separators.  Figure \ref{fig:PriorRatios}
shows $\log_{10}(\pi(\mathcal{G}) / \pi(\mathcal{G}'))) $ for different graphs $\mathcal{G}, \mathcal{G}'$.  Specifically, decreases in $b$ result in increased prior probability on models with few separators; in addition we see that as $a$ is increased, more mass is put on models with many cliques.  In contrast, we also plot the same ratio for the binomial prior (\ref{eq:binomial}).  From this one can see the limited control such a prior gives, favouring small models in terms of number of edges, but putting very little mass on models, for example, which feature clusters of fully-connected nodes (as in $\mathcal{G}_3$) and therefore have sparse covariance matrices.

Selecting the appropriate cohesion functions in equation (\ref{eq:cohesion}) is a difficult problem, but one for which we may gain insight from the existing literature on product partition models (\cite{Crowley1997, Quintana2003, Quintana2006}).  For instance, one may use Figure \ref{fig:PriorRatios} to select $a$ and $b$ to best fit with prior intuition regarding the features of the graph, then verify the choice through generation of Monte Carlo samples from the prior as in Figure \ref{fig:PriorSamples}.  Alternatively, cross-validation or related methods may be used to select $a$ and $b$; due to the potential computational cost of such methods, sequential Monte Carlo approaches may be used to speed up prior distribution selection (\cite{Bornn2010b}).

Given that the likelihood decomposes as (\ref{eq:likelihood}) and the prior is of the form (\ref{eq:cohesion}), the posterior will also be of the form (\ref{eq:cohesion}) with cohesions $\psi_{C}(C_{j})p(y_{C_{j}})$ and $\psi_{S}(S_{j})p(y_{S_{j}})$.  The prior admits several other attractive properties and connections with well-known clustering methods as well.  If $\psi_{S}(S_{j})\rightarrow\infty$ for all $S_{j}\neq\emptyset$, then Equation
(\ref{eq:cohesion}) reduces to the following model%
\begin{equation*}
\pi(\mathcal{G})\propto\prod_{j=1}^{n_c}\psi_{C}(C_{j})
\end{equation*}
if $n_s=0$ and $0$ otherwise. The resulting prior puts only positive mass on
graphs with no separators. It has been
introduced as a prior over partitions by \cite{Hartigan1990} and
\cite{Barry1992,Barry1993} under the name of
\textit{product partition models}. In the particular case of
(\ref{eq:cohesion}) with $b\rightarrow 0$, the prior over
$\mathcal{G}$ reduces to
\begin{equation*}
\pi(\mathcal{G})=\frac{a^{n_c}\Gamma(a)}{\Gamma(a+n)}\prod_{j=1}^{n_c}(|C_{j}|-1)!
\end{equation*}
As shown by \cite{Quintana2003} (see also \cite{Quintana2006}), this is the
distribution over partitions induced by a Dirichlet
Process~\cite{Ferguson1973,Antoniak1974}. We also have
\begin{equation*}
\mathbb{E}(n_c)=\sum_{i=0}^{n-1}\frac{a}{a+i}\simeq a
\log(1+n/a)+\gamma,\ \ \ \ \ var(n_c)=\sum_{i=1}^{n-1}\frac{a
i}{(a+i)^{2}}%
\end{equation*}
where $\gamma$ is Euler's constant and
\begin{equation*}
\text{pr}(n_c=k)=s(n,k)a^{k}\Gamma(a)/\Gamma(a+n)
\end{equation*}
where the coefficients $s(n,k)$ are the absolute values of Stirling numbers
of the first kind \cite{Antoniak1974}.  In this limiting case, the number of cliques increases logarithmically with the number of nodes.

\subsection{Extensions}
Motivated by the larger class of exchangeable partition functions~\cite{Pitman1995,Lau2007}, we can also consider four-parameters models, allowing more control over the relative sizes of the cliques/separators
\begin{equation*}
\pi(\mathcal{G})\propto \frac{\prod_{j=1}%
^{n_c}(a_2+a_1(j-1))\frac{\Gamma(|C_j|-a_1)}{\Gamma(1-a_1)}}{\prod_{j=1}^{n_s}(b_2+b_1(j-1))\frac{\Gamma(|S_j|-b_1)}{\Gamma(1-b_1)}}
\end{equation*}
where $a_2>-a_1,0\leq a_1<1$, likewise for $b_1,b_2$. The above model reduces to (\ref{eq:ppm_prior}) when $a_1=b_1=0$. We can also consider models that control the maximal number of cliques/separators
\begin{equation*}
\pi(\mathcal{G})\propto \frac{\prod_{j=1}%
^{n_c}(c_1-j+1)\frac{\Gamma(c_2+|C_j|)}{\Gamma(c_2)}}{\prod_{j=1}^{n_s}(d_1-j+1)\frac{\Gamma(d_2+|C_j|)}{\Gamma(d_2)}}
\end{equation*}
where $c_1,c_2,d_1,d_2>0$, and $c_1>d_1$ are the maximal number of cliques/separators. These two models respectively admit as limiting cases the distribution over partitions induced by the two-parameter Poisson-Dirichlet distribution and the finite Dirichlet-multinomial distribution, see e.g.~\cite{Lau2007} for further details on these distributions.  Using such extensions, one is able to both extend the product graphical model prior to control relative sizes and the maximal number of cliques and separators, as well as borrow from the wealth of literature on Dirichlet and related distributions to gain insight into the prior distribution's characteristics.


\section{Example: Modeling Agricultural Output of Different Species}

Determining agricultural policies to govern crop production, harvesting, and export is a challenge fraught with high variability both temporally and spatially.
Enabling effective crop management, handling, and marketing, thus requires accurate understanding of crop yield that account for and explain these variations.  While much effort has been made in developing models for predicting single crops (\cite{Stone2005a, Potgieter2006a}), little effort has been made in understanding statistically the relationship between crop yield of different crop varieties.

Understanding the connection between yields of different crop varieties is valuable for a multitude of reasons.  Firstly, because certain crops are planted and harvested at different times, the management of one crop might benefit from knowledge obtained from harvesting a similar crop earlier in the year.  Additionally, by accounting for correlation between different crops, insurers might better cover themselves against extreme events and better control insurance rates for farmers.  Lastly, farmers themselves might wish to ensure some level of stability in their income, and therefore might prefer to plant crops which are uncorrelated in yield.  Through such a practice, a farmer would be proactive in preventing disasters across his entire crop portfolio.  Simply by looking at the resulting undirected graph, a farmer could select two crops which do not have a path connecting them, and are therefore uncorrelated.

We examine the total production (in thousands of bushels) of $24$ crops in the state of California from the years $1990$ to $2009$ ($20$ years).  The data is compiled from the U.S Department of Agriculture website, where a considerable database is available for viewing and analysis.  The 24 crops include, for example, several varieties of wheat, rice, and beans.
We use the now-standard Gaussian hyper-inverse Wishart model: the likelihood of yield is given in (\ref{eq:likelihood}) and (\ref{eq:sublikelihood}), and the prior for the covariance matrix $\Sigma$ is hyper-inverse Wishart, which factorizes similarly to (\ref{eq:likelihood}), as a ratio of inverse Wishart distributions over cliques and separators (\cite{Giudici1996a}).  See \cite{Carvalho2009b} for some alternative marginal likelihoods based on fractional Bayes factors which can help to induce parsimony.  The parameters chosen for the hyper-inverse Wishart distribution are as described in \cite{Jones2005a}; we focus on the specification of $\pi(\mathcal{G})$.  Looking at the list of crops, one would expect that there will be clustering of the yields according to crop characteristics.  For instance, it would be reasonable to expect the yield of beans to be correlated with each other.  We also seek an interpretable graph, namely one with small complexity (in terms of number of edges and/or separators).  The first such prior we examine is the binomial prior of \cite{Jones2005a} with $\rho = 2/(n-1)$, chosen due to its prevalence in the literature.  While such a prior allows for penalization on the number of edges, no control is available over clustering.  In contrast, by using the prior (\ref{eq:ppm_prior}), we can set $b=.01$ to put strong penalization on the number of separators (and hence induce separation of the cliques and therefore sparsity in the correlation matrix), and set $a=.01$ to encourage a small number of cliques in the pursuit of simplicity in the resulting graph.

\begin{figure}
	 \centering
      \includegraphics[width=0.95\textwidth]{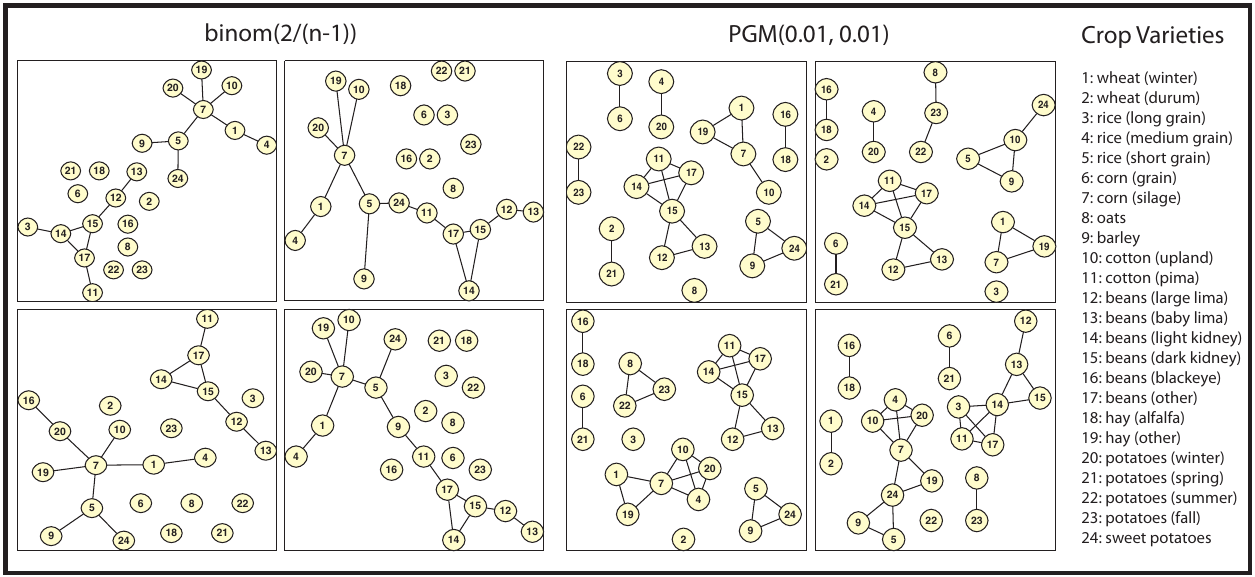}
  \caption{Four samples with highest posterior probability from crop yield model using binomial and product graphical model  (PGM) priors.  We see the bean yields (nodes 12 through 17) seem to cluster together, as do summer and fall potatoes (nodes 22 and 23).  We also observe that the product graphical model prior induces separated cliques, whereas the binomial prior results in long strings and trees of connected variables.  As a result, the product graphical model prior will induce sparsity in the resulting posterior covariance.}
	\label{fig:AgriPosteriorSamples}
\end{figure}

We run MCMC of length $10$ million over the space of decomposable graphs (\cite{Giudici1999a}) for both the binomial and product graphical model priors, thinning to every $100$ samples.  With both priors, one may save computational resources by making local moves, merging and splitting cliques within the Markov chain.  As a result, one need not re-determine the structure of the entire graph at each move.  Figure \ref{fig:AgriPosteriorSamples}
 shows the 4 graphs with highest posterior probabilities from each prior.  The product graphical model prior results in the top 4 graphs having posterior density values in the range 0.11 to 0.49, whereas for the binomial the range is 0.04 to 0.06, indicating that the binomial prior spreads mass much more evenly across distributions relative to the product graphical model prior with $a=b=0.01$.  Immediately evident from the figure is the different forms resulting from each prior.  Specifically, the binomial prior induces long strings of nodes with many separators, whereas the product graphical model posterior reflects our prior beliefs that variables will cluster together, resulting in sparsity in the correlations between variables.  A commercial farmer desiring to plant two plots with uncorrelated crops to minimize the risk of loss might reach quite different conclusions from each prior.  Specifically, the large strings of nodes from the Binomial prior suggest correlation between the majority of crops.  The farmer might not plant winter wheat (planted in late fall) and a strain of beans (harvested in early fall) on his two plots, despite their very different growing seasons, due to their connection in two of the highest posterior probability graphs in Figure \ref{fig:AgriPosteriorSamples}.  In contrast, the separation of cliques from the product graphical model prior (\ref{eq:ppm_prior}) would allow these crops to be planted together.  Such decisions could be made from the highest posterior graph, or by conducting Bayesian model averaging to obtain the expected utility of a given decision.

To gain an understanding of the product graphical model prior's prediction performance, we split the data into a training set (first $12$ years) and testing set (last $8$ years).  After simulating from the posterior distribution arising from the binomial and product graphical model priors, we use Bayesian model averaging via the marginal likelihood evaluated on the test data to judge the model's prediction performance.  We evaluate the resulting posterior predictive evaluated on the test set in Table \ref{tbl:PredictiveDensity}; indeed, the product graphical model prior provides better prediction in this example, even over a variety of parameter choices.  We also show the number of edges for each model, indicating that sparsity in terms of edges alone is not responsible for the improved prediction.
\begin{table}
	\centering
\caption{Log predictive density evaluated on test data using various priors}{
	\begin{tabular}{lrrrrrrrr}
		Distribution: & \hphantom{X} &\multicolumn{2}{c}{Binomial} & \hphantom{X} & \multicolumn{3}{c}{PGM}\\
		Parameters: & \hphantom{X} &$2/(24-1)$ & $0.5$ & \hphantom{X} &($0.01,0.01$) & ($0.1,0.1$) & ($1,1$) \\
		Avg. Log Predictive: & \hphantom{X} &$-688$ & $-707$ & \hphantom{X} &$-675$ & $-677$ & $-686$ \\
		Avg. Number of Edges: & \hphantom{X} &$17.8$ & $29.6$ & \hphantom{X} &$18.1$ & $16.4$ & $20.3$ \\
	\end{tabular}}
	\label{tbl:PredictiveDensity}
\end{table}

\section{Example: Modeling 20th Century American Voting Patterns}

In an effort to demonstrate the product graphical model prior in higher dimensions, we now turn to the modeling of American voting data by state.  For each federal election from 1904 to 1976, occurring every four years, we measure the proportion of votes for the republican party in each of the $50$ states (\cite{Carr2005a}).  Our goal is to model and visualize correlation in voting pattern changes over the last century.  Some immediate questions come to mind: ``Do certain states have an important role in determining election outcomes?'', ``Are there groups of states which vote together, operating independently from the US as a whole?''

\begin{figure}
  \begin{center}
	\mbox{
   	\subfigure[PGM($10,10^{-3}$): HPD Graph 1]{\includegraphics[width=0.45\textwidth]{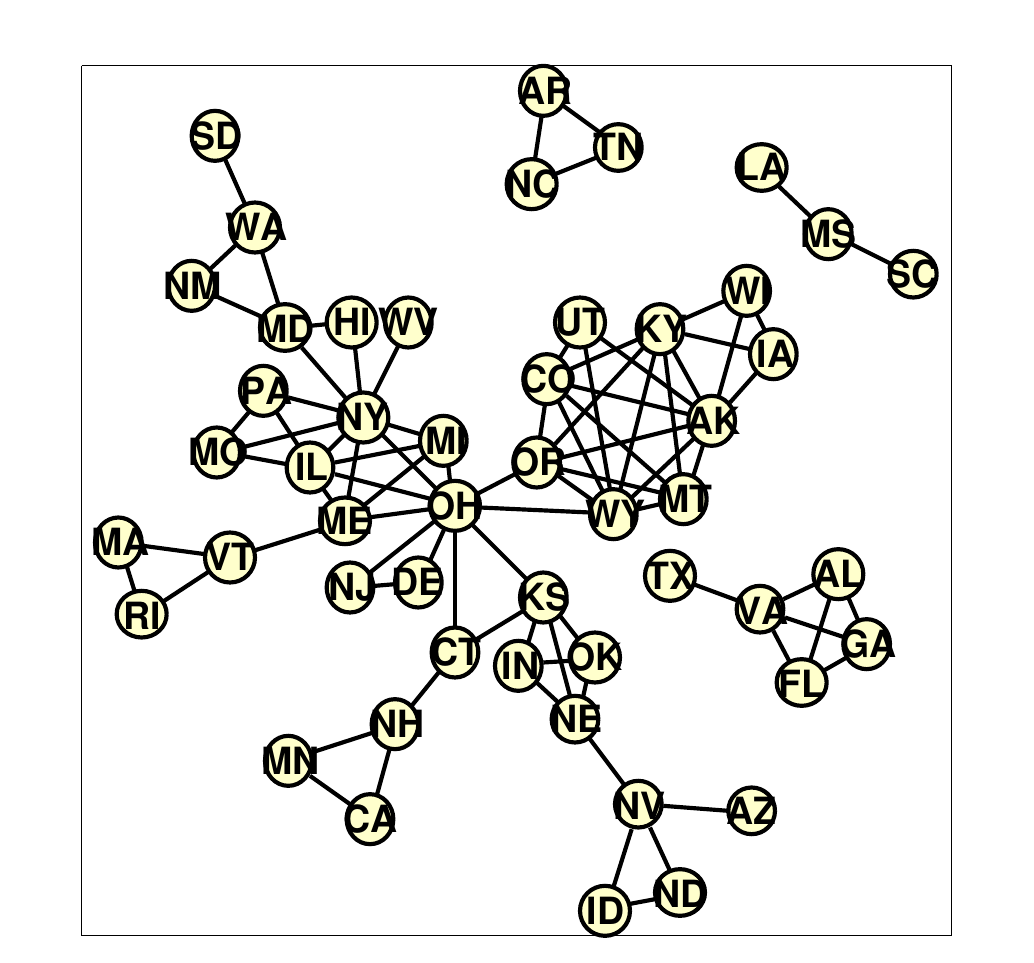}} \quad
      \subfigure[Binom(0.01): HPD Graph 1]{\includegraphics[width=0.45\textwidth]{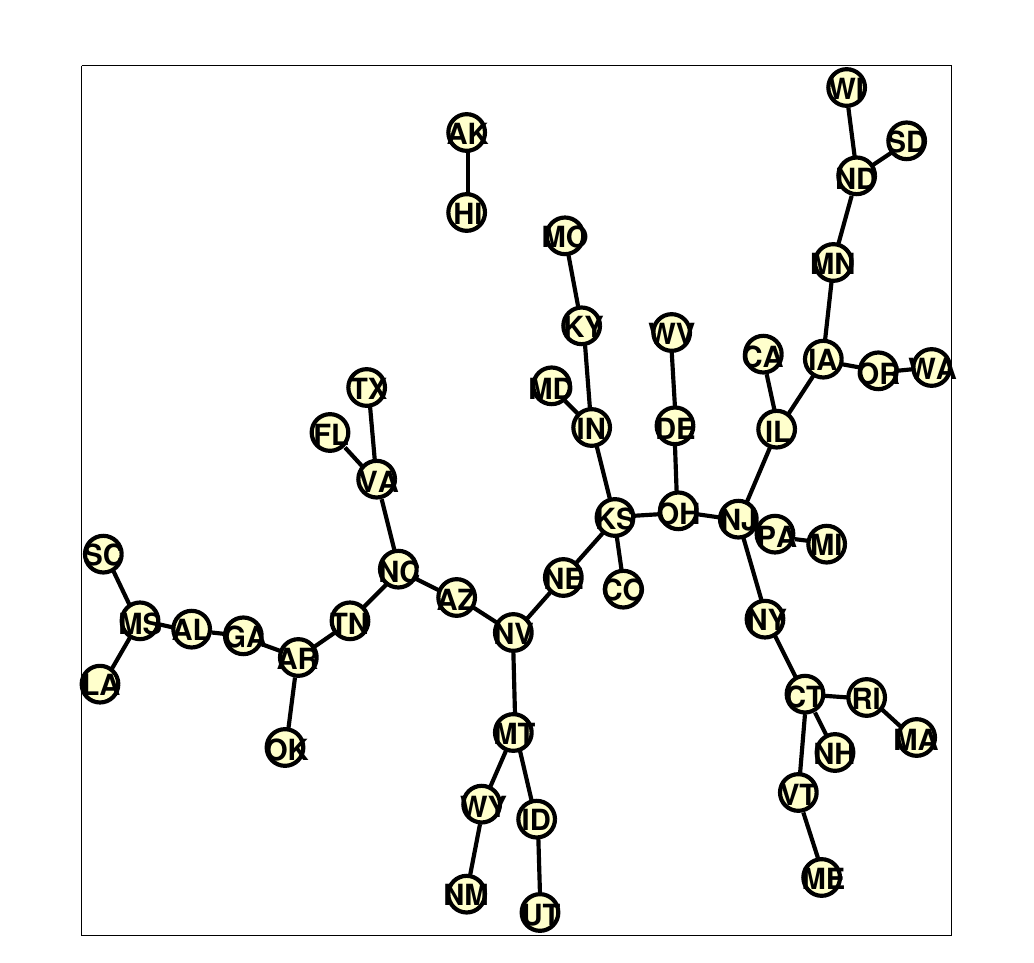}}
      }
    \mbox{
      \subfigure[PGM($10,10^{-3}$): HPD Graph 2]{\includegraphics[width=0.45\textwidth]{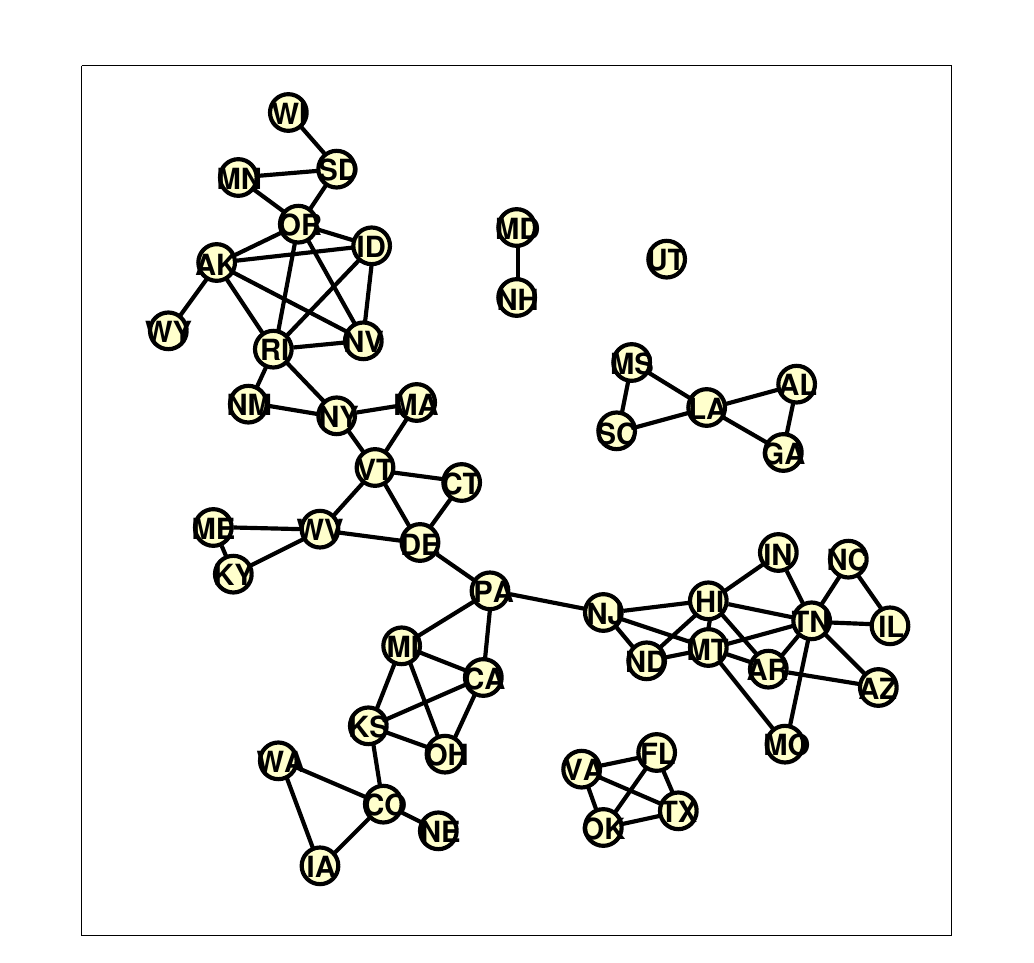}} \quad
      \subfigure[Binom(0.01): HPD Graph 2]{\includegraphics[width=0.45\textwidth]{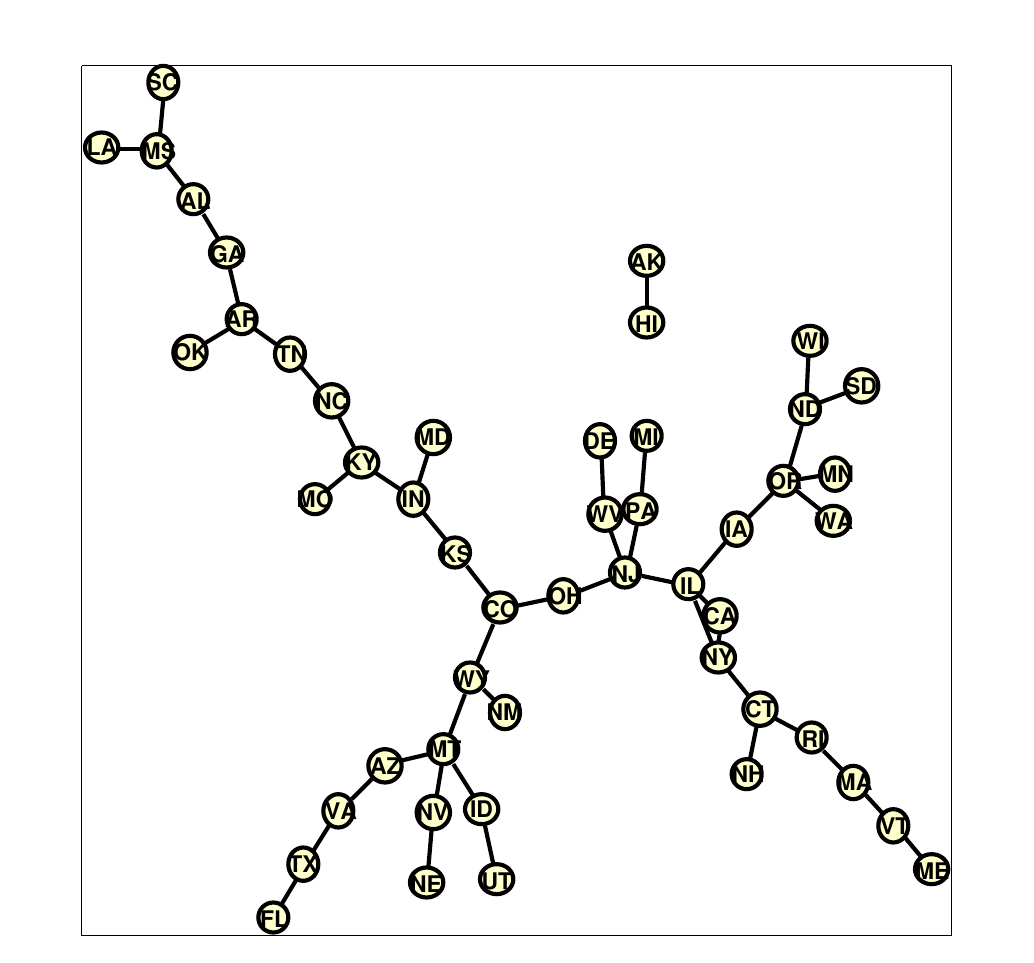}}

      }
    \caption{Voting example: two graphs with highest posterior density (HPD) from binomial and product graphical model priors.}
	\label{VotingHPD}
  \end{center}
\end{figure}

We proceed by exploring the posterior distribution resulting from the binomial prior with edge probability $0.1$, and the product graphical model prior with parameters $a=10, b=10^{-3}$, in an effort to make the overall number of edges resulting from each model comparable.  Figure \ref{VotingHPD} shows the two graphs with highest posterior density from each model.  As expected, the binomial graphs contain long strings of variables, while the product graphical model prior demonstrates clustering and grouping of variables.  While the binomial prior results in similar variables placed along the same string, the grouping from the product graphical model allows for clearer interpretation.  For instance, we immediately observe that the southern states (SC, MS, LA, AL, GA, TX, VA, FL) generally vote in a group.  Other patterns of interest also arise, including a close connection between AR, NC, and TN.  Also, notice that NY and KS are consistently the single node connecting clusters of variables.  As such, these states might be considered as key indicators of voting behavior.

\section{Discussion}


While we have focused on the Bayesian approach to covariance selection, significant work has also been done in a non-Bayesian framework.  A common approach involves placing an $\ell_1$ penalization on the precision matrix $\Sigma^{-1}$, which leads to sparse estimates (\cite{Meinshausen2006a,Yuan2007,Friedman2008}).  Closer to the heart of this paper, \cite{Marlin2009} examine the case of estimating $\mathcal{G}$ when clustering is expected, and therefore $\Sigma^{-1}$ exhibits block structure.  However, these models are neither decomposable nor generative.

While we have focused in this article on Gaussian graphical models, the prior defined in this article is far more general and can be used with any type of model for handling discrete or mixed data, see e.g. \cite{Madigan1995,Lauritzen1996}.  We have also considered the hyperparameters $a$ and $b$ to be known constants. Estimating them within the MCMC sampler would require one to compute the normalizing constant in \eqref{eq:ppm_prior}, which is in general not tractable. An exception of interest is the case $b\rightarrow 0$, where we can assign a gamma prior to $a$ and use the data augmentation algorithm described in~\cite{West1992} to update $a$ given the other variables.

In conclusion, the proposed product graphical model prior improves flexibility in modeling decomposable graphical models and borrows strength from the immense literature on product partition and related models.  The product graphical model prior allows one to encourage (or discourage) clustering of the graphs, and therefore can induce sparsity in the correlation matrix through clique separation; consequently, the product graphical model empowers practitioners to encapsulate their true prior beliefs to build a model more attuned to the problem at hand.

\singlespacing

\bibliographystyle{plain}
\bibliography{pgm,lukebornn}

\end{document}